\documentclass[prl,aps,groupedaddress,showpacs,twocolumn,superscriptaddress]{revtex4}
\usepackage{amssymb,amsmath,amsfonts,epsfig,graphicx,latexsym,bm,color,subfigure,braket}

\begin{document}
\title{Orbital Many-Body Dynamics of Bosons\\ in the Second Bloch Band of an Optical Lattice}

\author{J. Vargas}
\affiliation{Institut f\"ur Laserphysik, Universit\"at Hamburg, 22761 Hamburg, Germany}
\author{M. Nuske}
\affiliation{Institut f\"ur Laserphysik, Universit\"at Hamburg, 22761 Hamburg, Germany}
\affiliation{Zentrum f\"ur Optische Quantentechnologien, Universit\"at Hamburg, 22761 Hamburg, Germany}
\affiliation{The Hamburg Center for Ultrafast Imaging, Luruper Chaussee 149, Hamburg 22761, Germany}
\author{R. Eichberger}
\affiliation{Institut f\"ur Laserphysik, Universit\"at Hamburg, 22761 Hamburg, Germany}
\affiliation{Zentrum f\"ur Optische Quantentechnologien, Universit\"at Hamburg, 22761 Hamburg, Germany}
\author{C. Hippler}
\affiliation{Institut f\"ur Laserphysik, Universit\"at Hamburg, 22761 Hamburg, Germany}
\author{L. Mathey}
\affiliation{Institut f\"ur Laserphysik, Universit\"at Hamburg, 22761 Hamburg, Germany}
\affiliation{Zentrum f\"ur Optische Quantentechnologien, Universit\"at Hamburg, 22761 Hamburg, Germany}
\affiliation{The Hamburg Center for Ultrafast Imaging, Luruper Chaussee 149, Hamburg 22761, Germany}
\author{A. Hemmerich}
\affiliation{Institut f\"ur Laserphysik, Universit\"at Hamburg, 22761 Hamburg, Germany}
\affiliation{Zentrum f\"ur Optische Quantentechnologien, Universit\"at Hamburg, 22761 Hamburg, Germany}
\affiliation{The Hamburg Center for Ultrafast Imaging, Luruper Chaussee 149, Hamburg 22761, Germany}
\date{\today}

\begin{abstract}
We explore Josephson-like dynamics of a Bose-Einstein condensate of rubidium atoms in the second Bloch band of an optical square lattice providing a double well structure with two inequivalent, degenerate energy minima. This oscillation is a direct signature of the orbital changing collisions predicted to arise in this system in addition to the conventional on-site collisions. The observed oscillation frequency scales with the relative strength of these collisional interactions, which can be readily tuned via a distortion of the unit cell. The observations are compared to a quantum model of two single-particle modes and to a semi-classical multi-band tight-binding simulation of $12 \times 12$ tubular sites of the lattice. Both models reproduce the observed oscillatory dynamics and show the correct dependence of the oscillation frequency on the ratio between the strengths of the on-site and orbital changing collision processes.
\end{abstract}

\bibliographystyle{prsty}
\pacs{42.50.Nn, 06.30.Ft, 37.10.Jk, 37.30.+i} 

\maketitle
The ground state wave function of bosonic atoms in optical lattices \cite{Ver:92, Hem:93, Gry:01, Blo:05, Lew:07} can be generally chosen to be real and positive \cite{Fey:72, Wu:09, Li:16}, giving rise to a rather featureless physical scenario, in contrast to electronic condensed-matter lattice physics, where a more complex structure such as orbital degrees of freedom of higher bands typically provides a richer physical reality, for example in the case of transition metal oxides \cite{Tok:00, Mae:04}. In contrast to the well-controlled and comparatively simple platform of optical lattices, in electronic condensed-matter examples, however, the discrimination of physics related to orbital degrees of freedom from the multitude of other possible mechanisms is difficult. This has triggered growing interest to study atoms in metastable higher Bloch bands of optical lattice potentials \cite{Isa:05, Liu:06, Lar:09, Mue:07, Wu:09, Wir:11, Oel:13, Li:14, Koc:15, Koc:16, Li:16, Jin:19, Sha:20}. The presence of energetically degenerate orbitals with different angular momenta and orientations gives rise to multiple degenerate global band minima in the single-particle band structure at different high-symmetry points within the first Brillouin zone. This results in the intriguing consequence that even tiny energy scales as that of weak contact interactions play a decisive role in determining the structure of the lowest energy state in each band. The presence of degenerate local orbitals enables contact interaction processes, which change the orbital character of the colliding atoms \cite{Liu:06, Li:16, Hem:19}. Such processes have been identified as essential for the experimentally observed formation of multi-orbital Bose-Einstein condensates (BEC) with interaction-induced local angular momentum \cite{Liu:06, Oel:13, Koc:15, Li:16}. Theoretical proposals have also pointed out the possibility of global angular momentum \cite{Lib:16, Xu:16}. While equilibrium phases have been studied in some detail, the study of non-equilibrium scenarios has remained limited to one-dimensional examples \cite{Wan:16, Niu:18}.

In this work, we experimentally and theoretically explore quantum dynamics reminiscent of Josephson oscillations of bosonic atoms in the second band of an optical lattice, which provides a double well structure in quasi-momentum space with two inequivalent energy minima. These oscillations are driven by the interplay between orbital changing collisions and conventional on-site collisions. Their relative strength, which can be readily tuned via a distortion of the unit cell, determines the oscillation frequency. Note that orbital changing collisions show some analogy to spin changing collisions \cite{Wid:05}. Our work is the first to extend the experimental study of interaction dynamics in optical lattices with bosons from the familiar lowest band examples to orbital optical lattices, which possess higher order orbitals. The future perspective of this work is to trigger further research aiming at a better understanding of the nature of contact interaction in the presence of orbital degrees of freedom, including processes as the spontaneous generation of local or global angular momentum \cite{Li:16}.

\begin{figure}
\includegraphics[scale=0.44, angle=0, origin=c]{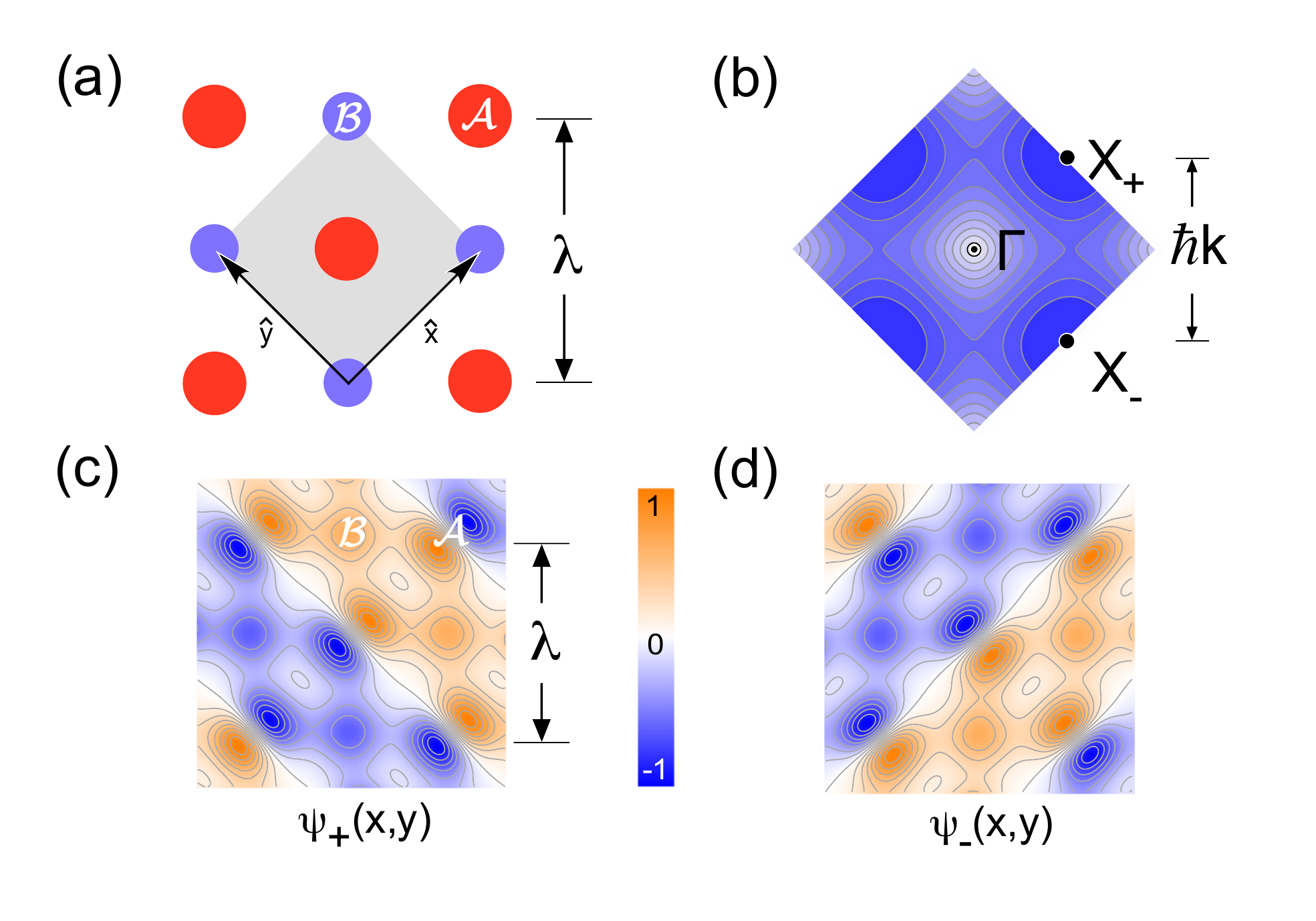}
\caption{(a) The bipartite lattice geometry with deep $\mathcal{A}$-sites and shallow $\mathcal{B}$-sites. The unit cell is shown by the gray rectangle. (b) The second Bloch band of the lattice in (a) is plotted across the first Brillouin zone with the two inequivalent energy minima at $X_{\pm}$ and the energy maximum at $\Gamma$ highlighted. Blue denotes low and white denotes high energy. (c) and (d) show contour plots of the Bloch functions $\psi_{\pm}$, corresponding to the $X_{\pm}$ points of the second band.}
\label{fig:Fig.1}
\end{figure}

We selectively populate one of the two degenerate global energy minima in the second Bloch band of an optical square lattice with a BEC and observe the subsequent dynamics, which displays a damped oscillation of population between both energy minima. This oscillation is exclusively driven by two kinds of collisional interactions, i.e., the on-site collisions of atoms in either of the three local orbitals $s$, $p_x$, and $p_y$, respectively, and an orbital changing collision mimicking a pair tunneling process \cite{Lia:09, Bad:09, Hem:19}, where two atoms colliding, e.g. in the  $p_x$ orbital at some lattice site, are both transferred to the $p_y$ orbital or vice versa. According to our model calculations, the frequency of the population oscillation scales with the relative strength of these collisional interactions, which can be readily tuned in the experiment. We implement two different models both describing essential aspects of our experimental findings: A quantum two-mode model involving the two Bloch modes associated with the energy minima of the second band, and a simulation of $12 \times 12$ tubular sites according to Fig.~\ref{fig:Fig.1}(a), treated in a four-band tight-binding approach accounting for nearest- and next-nearest neighbor tunneling and on-site contact interactions. 

We consider a two-dimensional (2D) optical square lattice in the $xy$-plane, composed of deep and shallow wells arranged according to the black and white fields of a chequerboard, see Fig.~\ref{fig:Fig.1}(a). In the third dimension, i.e. the $z$-direction, a nearly harmonic potential with $\Omega/2 \pi \approx 40\,$Hz is applied, such that the 3D lattice potential constitutes a 2D lattice of elongated sites. The 2D lattice potential is well approximated by $V(x,y) \, \approx  -V_{0}\, |\cos(kx) +  e^{i \theta} \cos(ky)|^2$ with $k = 2\pi/\lambda$ and $\lambda = 1064\,$nm. Adjustment of $\theta$ permits controlled rapid tuning of the potential energy difference $\Delta V \equiv  -4\, V_{0} \cos(\theta)$ between $\mathcal{A}$- and $\mathcal{B}$ wells. Technical details are given in Ref.~\cite{Koc:16}. In its second Bloch band this lattice provides two inequivalent degenerate local energy minima at the high-symmetry points $X_{+}$ and $X_{-}$, located at the edge of the first Brillouin zone (BZ), see Fig.~\ref{fig:Fig.1}(b). The corresponding Bloch functions, $\psi_{+}$ and $\psi_{-}$, composed of $p$ orbitals in the deep $\mathcal{A}$ wells and $s$ orbitals in the shallow $\mathcal{B}$ wells, are orthogonal real-valued standing waves (cf. Figs.~\ref{fig:Fig.1}(c) and (d)).

\begin{figure}
\includegraphics[scale=0.4, angle=0, origin=c]{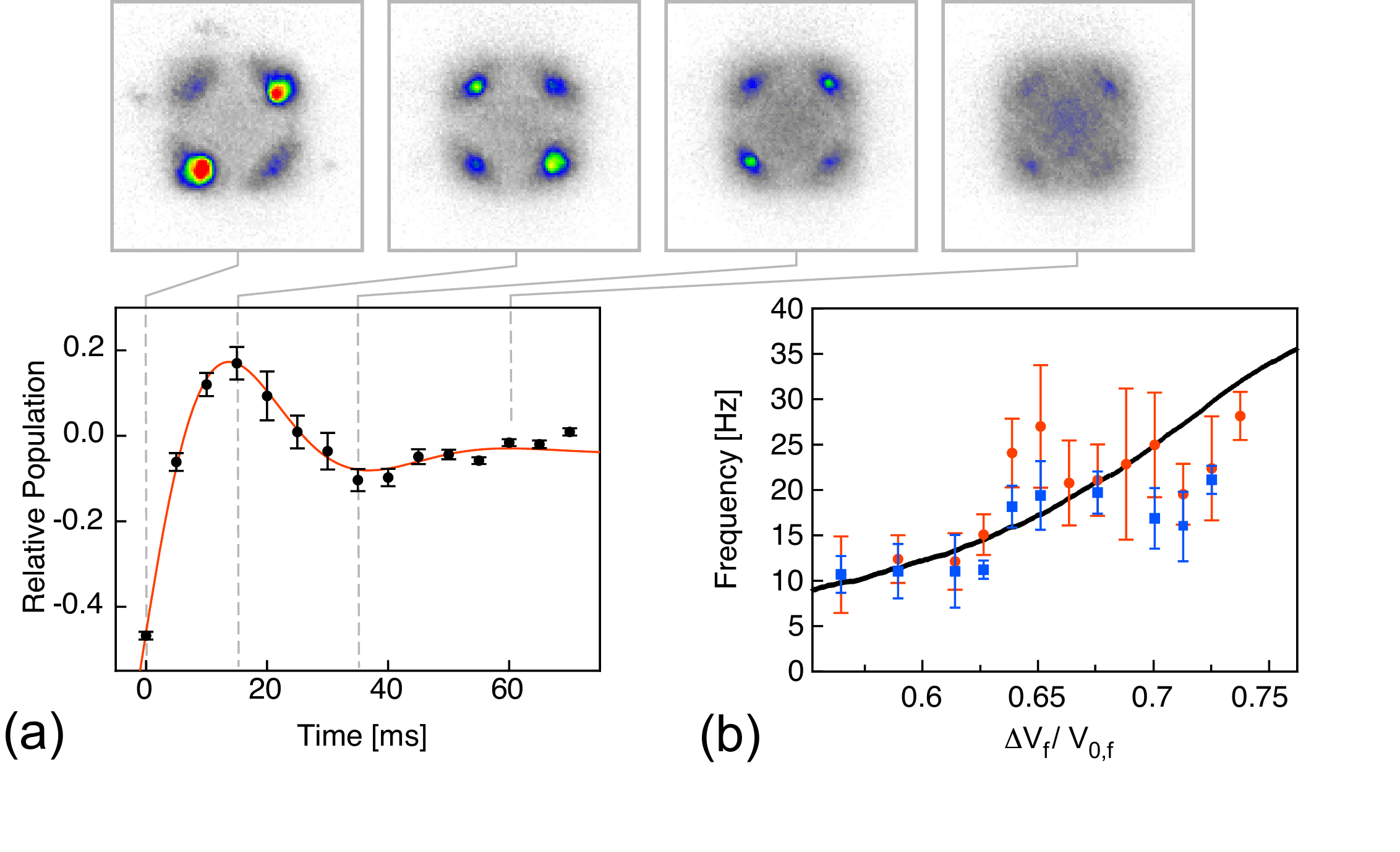}
\caption{(a) The temporal evolution of the relative population difference $\frac{n_{-} - n_{+}}{n_{-} + n_{+}}$ is shown for fixed $\Delta V_{f} = 0.725 \times \,V_{0,f}$ after initially a BEC is formed at the $X_{+}$ point. The red solid line is a fit by an exponentially damped harmonic oscillation. The error bars show the standard deviations of the mean for a set of ten measurements. (b) The observed oscillation frequencies obtained from fitting data as in (a) are plotted versus $\Delta V_{f}$. The errors show the standard deviations found in the fits. The red disks and the blue squares represent two measurement series evaluated via momentum spectra and band mapping, respectively. The solid black line shows a calculation using a two-mode model.}
\label{fig:Fig.2}
\end{figure}

We prepare the initial state with the following protocol. A nearly pure rubidium BEC with up to $6 \times 10^4$ atoms in the $\ket{F=2, m_F=2}$ hyperfine state at about $50\,$nK temperature is initially loaded at the $\Gamma\,$point in the lowest band of the lattice potential with $\Delta V_{i} = -1.23\, V_{0,i}$ and $V_{0,i} = 4.3 \,E_{\textrm{rec}}$. Here, $E_{\textrm{rec}}\equiv \hbar^2 k^2 / (2 m)$ denotes the single-photon recoil energy and $m$ is the atomic mass. At this stage, the negative sign of $\Delta V_{i}$ indicates that the $\mathcal{B}$-wells are deep and the $\mathcal{A}$ wells are shallow. The atoms reside nearly exclusively in the $\mathcal{B}$ wells. Next, a magnetic field gradient realizes a magnetic force. The gradient is applied for $0.65\,$ms, such that the atoms undergo a half-cycle of a Bloch oscillation, and hence are transferred to the $X_{+}$ point. Details of this step are deferred to Ref.~\cite{Sup}. Finally, $\Delta V$ and $V_0$ are ramped up in $0.3\,$ms to final values $\Delta V_{f} \in [0.55, 0.75] \times \,V_{0,f}$ and $V_{0,f} = 7.2 \,E_{\textrm{rec}}$. According to the positive value of $\Delta V_{f}$ the roles of $\mathcal{A}$- and $\mathcal{B}$ wells are swapped such that a condensate at the $X_{+}$ point in the second band is formed with $25.000 \pm 2.000$ atoms \cite{Koc:16, Sup}. Subsequently, the temporal evolution of the relative population difference between the $X_{-}$- and the $X_{+}$ point $\frac{n_{-}-n_{+}}{n_{-}+n_{+}}$ is recorded. This quantity is retrieved by performing band mapping or alternatively, by recording momentum spectra, and counting the atoms in the vicinity of the $X_{\pm}$ points. Details are found in Ref.~\cite{Sup}. An example for $\Delta V_{f} = 0.725 \times  \,V_{0,f}$ is shown in Fig.~\ref{fig:Fig.2}(a). A strongly damped oscillation is observed at a frequency of $21.6\,$Hz. Corresponding band mapping pictures recorded at times indicated by dashed gray lines are shown on the upper edge of Fig.~\ref{fig:Fig.2}(a). The red solid line is a fit with a single exponentially damped harmonic oscillation. In Fig.~\ref{fig:Fig.2}(b), for each data point the procedure to obtain Fig.~\ref{fig:Fig.2}(a) is repeated and the observed oscillation frequencies $\nu_{\textrm{osc}}$ are plotted versus $\Delta V_{f}$. The plot shows an increase of $\nu_{\textrm{osc}}$ with increasing $\Delta V_{f}$. 

In the following, we will compare the data points in Fig.~\ref{fig:Fig.2}(b) with two distinct models, which will both allow us to directly connect the experimental parameter $\Delta V_{f}$ with the amplitude ratio between an orbital interaction process exchanging pairs of atoms between $p_x$ and $p_y$ orbitals and conventional Hubbard-like on-site interaction. We begin with a minimal model of the two Bloch modes $\psi_{+}$ and $\psi_{-}$. According to Ref.~\cite{Hem:19}, the Hamiltonian reads
\begin{eqnarray} 
\label{eq:Hamiltonian}
H &=&  \frac{g_0}{2} \left[\hat{n}_{+} (\hat{n}_{+}-1) + \hat{n}_{-} (\hat{n}_{-} - 1)\right]  \\ \nonumber
&+& \frac{g_1}{2} \left[4\, \hat{n}_{+} \hat{n}_{-}  +  \hat{ \psi}_{+}^{\dagger} \hat{ \psi}_{+}^{\dagger} \hat{ \psi}_{-} \hat{ \psi}_{-} + \hat{ \psi}_{-}^{\dagger} \hat{ \psi}_{-}^{\dagger} \hat{ \psi}_{+} \hat{ \psi}_{+}     \right]
\end{eqnarray}
with $\hat{ \psi}_{\pm}$ denoting the annihilation operator for the Bloch modes $\psi_{\pm}$ and $\hat{n}_{\pm} \equiv \hat{\psi}_{\pm}^{\dagger} \hat{ \psi}_{\pm}$ the corresponding number operators. As seen in Eq.~(\ref{eq:Hamiltonian}), $g_0$ corresponds to a conventional Hubbard on-site interaction, while the expression controlled by $g_1$ contains a pair exchange term between both modes, which changes the orbital flavor. As detailed in Ref.~\cite{Sup}, the collision parameters $g_0$ and $g_1$ can be expressed as $g_{0} = g_{\textrm{2D}}\, I_{0}(\Delta V_f)$ and $g_{1} = g_{\textrm{2D}} \,I_{1}(\Delta V_f)$ with an effective 2D collision energy $g_{\textrm{2D}}$ and the dimensionless integrals $I_{0}(\Delta V_f) \equiv A\, \int_{A} dx dy\, |\tilde{\psi}_{\pm}| ^4$ and $I_{1}(\Delta V_f) \equiv A \,\int_{A} dx dy\, |\tilde{\psi}_{+}| ^2 |\tilde{\psi}_{-}|^2$, where $\tilde{\psi}_{\pm}$ denote the Bloch wave functions normalized to a single unit cell of the lattice with area $A = \lambda^2 / 2$. In order to model the observation in Fig.~\ref{fig:Fig.2}(b), the following steps are performed. A numerical band calculation is performed to obtain $\tilde{\psi}_{\pm}$ and hence the integrals $I_{0}(\Delta V_f)$ and $I_{1}(\Delta V_f)$ as functions of $\Delta V_{f}$. The effective collision parameter $g_{\textrm{2D}}$ is expressed in terms of the 3D collision parameter for rubidium atoms (c.f. Ref.~\cite{Sup}). The Schr{\"o}dinger equation for the Hamiltonian of Eq.~(\ref{eq:Hamiltonian}) is solved for $N = 5.000$ particles, with the initial condition that all atoms reside at $X_{+}$. This leads to the full time evolution of the system state $\ket{\Psi(t)}$. Finally, the expectation value $\bra{\Psi(t)} (\hat{n}_{+} - \hat{n}_{-}) \ket{\Psi(t)} / N $ is obtained, a fast Fourier transform of this quantity is calculated and the frequency $\nu_{\textrm{osc}}$ of the dominant spectral component is determined. To estimate the prediction for $N=24.000$ particles (which results in a good match with the observations), we calculate $\nu_{\textrm{osc}}$ as a function of $N$, for the range of $N=20$ to $N=500$. The resulting dependence of $\nu_{\textrm{osc}}$ on $N$ is described by a power-law of the form $\sim N^{0.8896}$, within a relative error of $10^{-4}$. Utilizing this dependence, we extrapolate the value of $\nu_{\textrm{osc}}$ calculated for $N=5.000$ to find that for $N=24.000$ (c.f.~\cite{Extrap}). The resulting $\nu_{\textrm{osc}}$ plotted against $\Delta V_{f}$ is shown as the black solid line in Fig.~\ref{fig:Fig.2}(b). A detailed treatment of the underlying two-mode model is found in Ref.~\cite{Hem:19}. Remarkably, the observed ascending trend of $\nu_{\textrm{osc}}$ is well reproduced by the model although band relaxation, heating, and particle loss are neglected here. The band calculation of $I_{0}(\Delta V_f)$ and $I_{1}(\Delta V_f)$ shows that within the accessible range $\Delta V_f\in[0.2,0.73]\times V_{0,f}$, increasing $\Delta V_{f}$ acts to increase the quantity $g_2 \equiv 1- g_1/g_0$. The two-mode model \cite{Hem:19} predicts self-trapping (c.f. Ref.~\cite{Alb:05}) to occur if $g_2 > 2/3$, which, however, corresponds to values of $\Delta V_{f}$ well outside of this range, such that the band structure would no longer support a stable BEC in the second band.

\begin{figure}
\includegraphics[width=\linewidth]{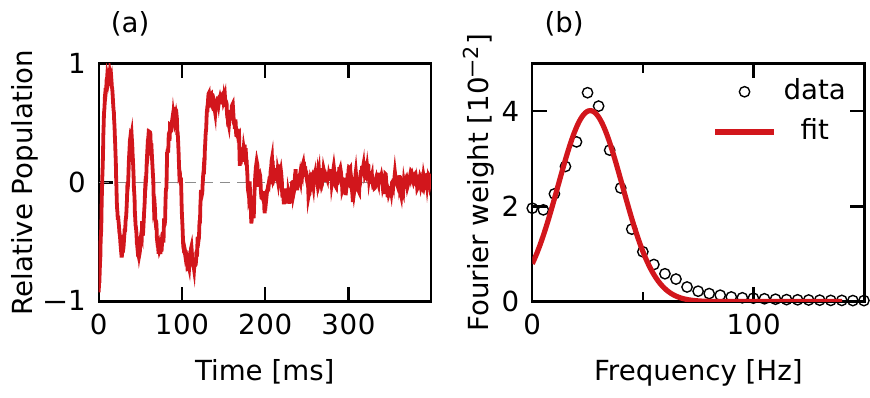}
\caption{(a) Relative population difference of $X$ points $\frac{n_{-} - n_{+}}{n_{-} + n_{+}}$ according to our classical-field-theory simulations at $\Delta V_f = 0.69\,V_0$. The panel shows the oscillation for a single random initialization. (b) Fourier spectrum of the oscillation shown in (a) averaged over 500 random initializations, see black circles. The peak of the Fourier transform determines the main oscillation frequency at the given value of $\Delta V_f$. The red line shows a Gaussian fit to the Fourier spectrum. For details on the fitting routine see Ref.~\cite{Sup}. For both panels the temperature is $T=76.8\,{\rm nK} \approx 0.8\,E_{\rm rec}/k_B$ and hence similar to the experimental temperature.}
\label{fig:Fig.3}
\end{figure}

\begin{figure}
\includegraphics[width=\linewidth]{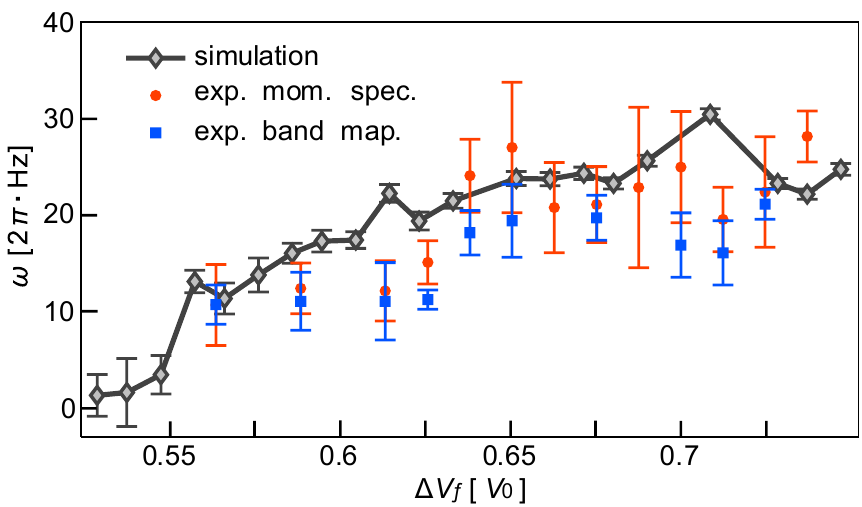}
\caption{Dominant oscillation frequencies obtained from c-field simulations, see black diamonds, are compared to experimental data, shown by blue squares and red disks, which are repeated from Fig.~\ref{fig:Fig.2}(b). For each $\Delta V_f$ we determine the averaged Fourier spectrum, as exemplarily shown in Fig.~\ref{fig:Fig.3}(b) and plot the dominant frequency obtained from a Gaussian fit. The errors in the determination of the positions of the maxima of the fitted Gaussians are mostly smaller than the data symbols.}
\label{fig:Fig.4}
\end{figure}

In order to obtain a more realistic description including dissipation and loss, we turn to a model consisting of $12 \times 12$ tubular sites according to Fig.~\ref{fig:Fig.1}(a) treated in a tight-binding approach accounting for nearest-neighbor and next-nearest neighbor tunneling and the same on-site collisions already included in the model in Eq.~(\ref{eq:Hamiltonian}). The tubes are mapped onto a 1D lattice by discretizing them in real space with a discretization length of $0.13\,\mu$m. We simulate the experimental loading protocol and the subsequent dynamics using classical-field-theory techniques, see Ref.~\cite{Sup}. We initialize the c-field propagation from a thermal ensemble of temperature $T$ using Monte Carlo sampling with parameters $V_{0,i} = 4.3\,E_{\rm rec}$ and $\Delta V_i = -1.23\, V_{0,i}$. We transfer the atoms to the $X_{+}$ point using phase-imprinting, see Ref.~\cite{Sup}, and quench the potential offset to its final value $\Delta V_f\in[0.2,0.73]\times V_{0,f}$ with $V_{0,f}=7.2\, E_{\rm rec}$. The resulting time evolution of the relative population of the $X$ points is shown in Fig.~\ref{fig:Fig.3}(a). For a single initialization and a temperature $k_B\,T= 0.8\,E_{\rm rec}$, similar to what is realized experimentally, we observe coherent oscillations between the $X_{+}$ and $X_{-}$ points during the first hundred milliseconds before damping by decay to the lower band sets in. In Ref.~\cite{Sup} we show that at lower temperatures, band relaxation and decay of the condensate fractions become negligible, and hence, for single initializations, coherent oscillations prevail for very long times. However, the frequency and phase of these oscillations vary for different initializations, which leads to additional strong damping via decoherence, when averaging over multiple initializations is performed (See Ref.~\cite{Sup}). Recall that the experimental data in Fig.~\ref{fig:Fig.2}(a) correspond to an average over ten initializations. As illustrated by the error bars, the error for a single initialization increases during the first $30\,$ms, thus reflecting this expected decoherence. Only for later times, when atom loss in the second band sets in, the error bars decrease again. For our numerical simulations we estimate the dominant oscillation frequencies by calculating the power spectrum. We Fourier transform the relative population of the $X$ points for each random initialization and subsequently average the Fourier spectra. The result is shown in Fig.~\ref{fig:Fig.3}(b). As expected, we find a broad Fourier peak, where many different oscillation frequencies contribute. We fit a Gaussian to the Fourier spectrum in order to extract the dominant oscillation frequency, for details see Ref.~\cite{Sup}. Figure \ref{fig:Fig.4} plots the results for different final potential offsets $\Delta V_f$ showing notable agreement with the experimental data repeated from Fig.~\ref{fig:Fig.2}(b). 

\begin{figure}
\includegraphics[width=\linewidth]{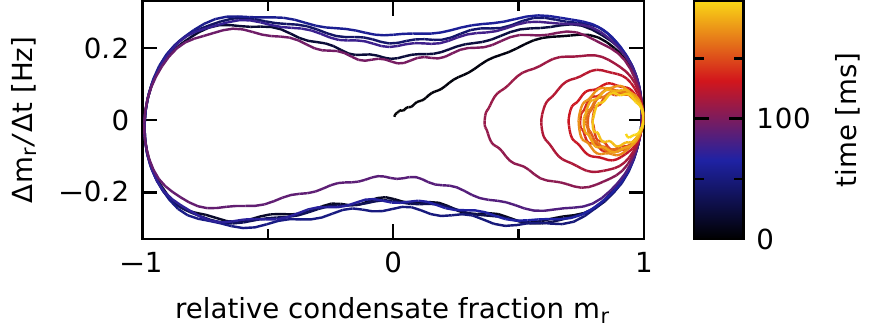}
\caption{Phase-space diagram of the relative population in the two lowest-energy many-body states $m_r$ for a single initialization of our simulations. Here,  $m_r=m_+(t)-m_-(t)$ and $m_{\pm}(t)=|\langle \Psi_{\pm}|\psi(t)\rangle|^2/m_{0}$. Furthermore $\psi(t)$ is the wave function obtained within our simulations and $m_{0}=|\langle \Psi_{+}|\psi(t)\rangle|^2+|\langle \Psi_{-}|\psi(t)\rangle|^2$ is the total number of condensed atoms in the second band. We consider an idealized case at lower temperature $T=0.5\,$nK and 4 times stronger interactions as compared to Fig.~\ref{fig:Fig.4}. We also use a different quench protocol that keeps $V_{0}=7\,E_{\rm rec}$ throughout the quench and changes $\Delta V$ from $\Delta V_i=-0.6\, V_0$ to $\Delta V_f=0.35 \,V_0$.}
\label{fig:Fig.5}
\end{figure}

Finally, we consider the equilibration dynamics after the quench within our c-field simulations. The two degenerate lowest-energy many-body states approximately form an $N$-fold occupation of either of the coherent superpositions $\Psi_{\pm} = \psi_{+} \, \pm i \, \psi_{-}$ of the two degenerate single-particle states $\psi_\pm$ \cite{Liu:06, Oel:13, Koc:16}. Their relative phases $\pm i$ minimize the energy associated with the pair exchange processes. We may consider the oscillatory dynamics that emerges in both experiment and simulations in terms of $\Psi_{\pm}$. To this end, we show in Fig.~\ref{fig:Fig.5} the normalized projection of the state $\psi(t)$ onto $\Psi_{\pm}$. We consider an idealized only weakly damped case by choosing a low temperature of $0.5\,$nK. Initially the atoms are prepared to occupy one of the $X$ points and hence their overlap with both $\Psi_{+}$ and $\Psi_{-}$ is $50\%$. Before damping sets in, the evolution is characterized by instanton-type dynamics \cite{Raj:82}, where the atoms perform perfect oscillations between $\Psi_{\pm}$. This reproduces the oscillations shown in Fig.~\ref{fig:Fig.3}(a) in the $\Psi_{\pm}$ basis, however at much lower temperature and hence lower damping. At the zero-crossings in Fig.~\ref{fig:Fig.3}(a), the atoms have unit overlap with one of the two many-body states $\Psi_{\pm}$. A typical single-implementation trajectory plotted in Fig.~\ref{fig:Fig.5} shows that this overlap alternates between $\Psi_{+}$ and $\Psi_{-}$, which amounts to an oscillating chirality. The slight inward shift of the trajectory at $50\%$ mixture is a result of the slow-down due to the free-energy barrier that separates the two lowest-energy states $\Psi_{\pm}$. Eventually, due to damping, the atoms do not have enough energy to cross this barrier and spontaneously pick either of the two states $\Psi_{\pm}$. In the subsequent second part of the dynamics the atoms perform damped harmonic oscillations in the corresponding free-energy minimum and hence have an overlap between $50\%$ and unity with this state. This example provides a limiting case of the many-dynamics of this system, for which the experimental results provide a more strongly damped realization.

In summary, we have studied quantum dynamics of a BEC in the second band of an optical lattice arising from the competition between orbital changing and conventional on-site collisions. A minimal quantum model and a more realistic model based on classical field simulations show quantitative agreement with the observations. For simulations of an idealized low-temperature scenario, we find coherent instanton-type dynamics characterized by oscillations between the two degenerate lowest-energy many-body states in the second band. Our work pioneers the exploration of quantum dynamics in optical lattices with orbital degrees of freedom and orbital degeneracies, which allows one to emulate a physical reality beyond $s$-band Hubbard physics.  

\begin{acknowledgments}
We thank Max Hachmann for his contributions in the initial stage of the experiment. We acknowledge partial support from the Deutsche Forschungsgemeinschaft (DFG) through the collaborative research center SFB 925 (M.N. and L.M.), the Cluster of Excellence CUI: Advanced Imaging of Matter - EXC 2056 - project ID 390715994 (M.N. and L.M.), and the individual grants program DFG-He2334/17-1 (A.H.). J.V.~is grateful to the National Agency for Research and Development (ANID) of Chile and its Ph.D.~scholarship program. M.N.~acknowledges support from Stiftung der Deutschen Wirtschaft. We thank Juliette Simonet, Klaus Sengstock and their entire team for useful discussions.
\end{acknowledgments}

\section{Supplemental Material}

\begin{appendix}

\section{Experimental procedures}
\begin{figure}[tb]
\centering
\includegraphics[width=0.5\textwidth]{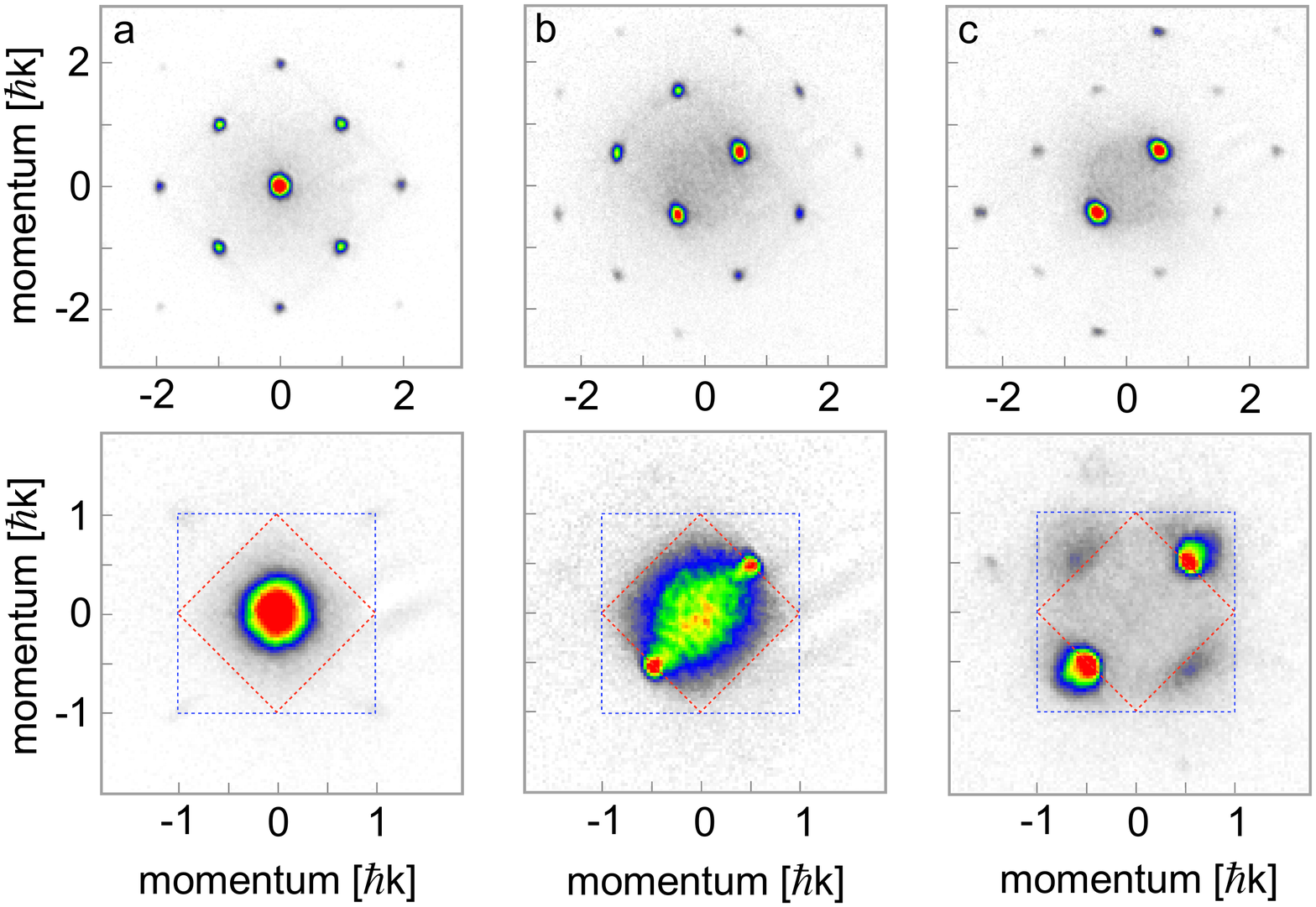}
\caption{Momentum spectra (upper panels) and corresponding band mapping images (lower panels) observed subsequent to the three steps of the quench protocol applied: (a) After the atoms are loaded into the $\Gamma-$point of the lowest Bloch band; (b) After the atoms are transferred to the $X_{+}$-point in the lowest band via half of a Bloch oscillation cycle; (c) After the atoms are transferred to the $X_{+}$-point in the second band. The dotted red and blue rectangles in the lower panels mark the boundaries of the first (region within the inner rectangle tilted by $45^\circ$) and second (region between outer non-tilted and inner tilted rectangle) Brillouin zones.}
\label{fig:Fig.A1}
\end{figure}
In this section, a detailed description of the experimental protocol is given, which consists of a quench composed of three steps and a subsequent observation of the induced dynamics. We start with a Bose-Einstein condensate (BEC) of $6 \times 10^4$ $^{87}$Rb atoms in the $\ket{F=2,m=2}$ hyperfine state in an isotropic magnetic trap with trap frequencies $(\omega_{x},\omega_{y},\omega_{z})= 2\pi \times (39, 42, 35)\,$Hz. We employ the bipartite square optical lattice potential, described in the main text, which provides tunability of the overall lattice depth $V_{0}$ and the potential energy difference $\Delta V$ between the $\mathcal{A}$ and $\mathcal{B}$-wells of the lattice. Tuning of $\Delta V$ is accomplished via control of the phase angle $\theta$ according to $\Delta V = -4 V_{0} \cos(\theta)$ (see main text). This angle corresponds to the path length difference in a Michelson interferometer set-up, used to implement the lattice potential, which can be servo-controlled with $\pi/300$ precision \cite{Hem:91, Koc:16}.   

In a first step, the lattice potential is ramped up in $100\,$ms to an initial value of the lattice depth $V_{0} = V_{0,i} \equiv 4.3 \,E_{\textrm{rec}}$ with fixed $\Delta V = \Delta V_{i} \equiv -1.23\, V_{0,i}$. Here, $E_{\textrm{rec}}$ denotes the single-photon recoil energy. As a result the BEC now resides in the lowest Bloch band of the lattice potential at the center of the first Brillouin zone (BZ), denoted as $\Gamma-$point. This is confirmed in Fig.~\ref{fig:Fig.A1}(a) by a momentum spectrum (upper panel) and a corresponding mapping of quasi-momentum space (lower panel). At this stage, the negative sign of $\Delta V_{i}$ indicates that the $\mathcal{B}$-wells are deep and the $\mathcal{A}$-wells are shallow. The atoms are located nearly exclusively in the $\mathcal{B}$-wells. 

In a second step after a waiting time of $10\,$ms, the BEC is transferred in $0.65\,{\rm ms}$ from the $\Gamma-$point to the $X_{+}$-point at the edge of the first BZ such that the atoms remain in the lowest band. This is accomplished by applying a constant magnetic field gradient that exerts a force such that the atoms undergo a half-cycle of a Bloch oscillation. The result is seen in Fig.~\ref{fig:Fig.A1}(b), which shows a significant population of the $X_{+}$-point. Collisional contact interaction between atoms with opposite momenta at the two opposite edges between the first and second BZ leads to a broad background of atoms scattered across the entire first BZ.

The final step is applied to transfer the BEC into the $X_{+}$-point of the second band. To this end, in $300\,{\rm \mu s}$ the lattice depth $V_{0}$ is ramped up to $V_{0,f} \equiv 7.2\,{\rm E_{rec}}$, while simultaneously $\Delta V$ is tuned to a final value $\Delta V_{f} \in [4.1,5.3]\,{\rm E_{rec}}$. According to the positive value of $\Delta V_{f}$ the roles of $\mathcal{A}$- and $\mathcal{B}$-wells are swapped such that the atoms now form a condensate at the $X_{+}$-point in the second band. This result is confirmed by a momentum spectrum and a band mapping plot in Fig.~\ref{fig:Fig.A1}(c).

After loading the lattice, performing a Bloch oscillation, and conducting a quench to load the second band, we end up with $25.000 \pm 2.000$ atoms at the $X_{+}$-point of the second band. Slightly more than half of the atoms of the initial BEC are lost to higher bands and a non-condensed fraction of atoms in the second band. 

\begin{figure*}[t!b]
\includegraphics[width=422pt]{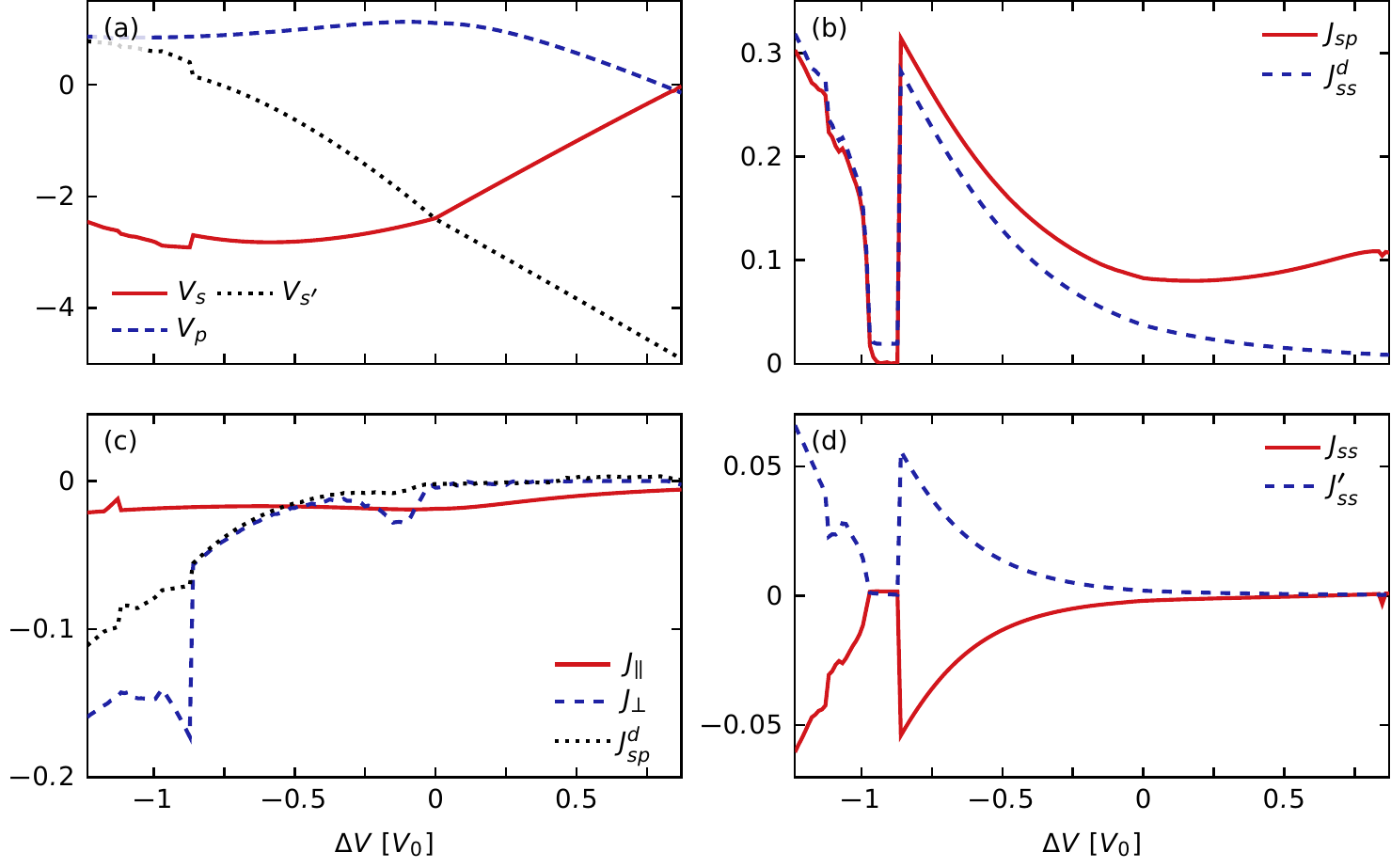}
\caption{Tight-binding parameters as a function of $\theta$ for $\lambda_L=1064\,{\rm nm}$ and the rubidium mass $m=87m_p$ in units of $E_{\rm rec}$. The lattice lattice depth at $\Delta V=-1.2\, E_{\rm rec}$ is $V_{0,i}=4.3\,E_{\rm rec}$ and is then linearly changed to $V_{0}=7.2\,E_{\rm rec}$ at $\Delta V=0$. For values $\Delta V\geq 0$ it is kept fixed at $V_{0}=7.2\,E_{\rm rec}$.}
\label{fig:Fig.A2}
\end{figure*}

To investigate the dynamics, we let the system evolve during a variable holding time up to $100\,{\rm ms}$ and observe the atoms via momentum spectra and band mapping images, analogous to those in Fig.~\ref{fig:Fig.A1}. For the band mapping technique, the lattice depth $V_0$ is adiabatically ramped down in $1.5\,$ms followed by a ballistic expansion during $30\,{\rm ms}$. For the case of momentum spectra, the lattice and trap potentials are switched off instantaneously ($< 1\,\mu$s) before the $30\,{\rm ms}$ ballistic expansion. Finally, the temporal evolution of the relative population difference between the $X_{-}$- and the $X_{+}$-point $\frac{n_{-} - n_{+}}{n_{-} + n_{+}}$ is recorded. This quantity is retrieved by counting all atoms in disk-shaped regions of interest (ROI) around the $X_{\pm}$-points and subtracting the atoms within ring-shaped ROIs of equal area enclosing the disk-shaped ROIs. The time dependence of the resulting relative populations, as exemplified in Fig.2(a) of the main text for $\Delta V/V_{0,f} = 0.725$, is fitted with a single exponentially damped harmonic oscillation $A \sin(2 \pi \nu_{\textrm{osc}}t) e^{-t/\tau}$. The frequencies $\nu_{\textrm{osc}}$ thus determined are plotted versus $\Delta V/V_{0,f}$ in Fig.2(b) of the main text.

\section*{Estimation of collision parameters}
Upon the simplifying assumption that the lattice in the $xy$-plane extends over $M$ unit cells of area $A$, the collision parameters in the two-mode Hamiltonian can be written
\begin{align*} 
g_0 &=  \int_{-\infty}^{\infty} dz \int_{M \times A} d^3 r \,|\psi_{\pm}(x,y)|^4 |\chi(z)|^4   \, , \\
g_1 &= \int_{-\infty}^{\infty} dz\int_{M \times A} d^3 r \,|\psi_{+}(x,y)|^2 |\psi_{-}(x,y)|^2 |\chi(z)|^4  \, ,
\end{align*}
where $\psi_{\pm}(x,y)$ denote the 2D Bloch functions associated with the $X_{\pm}$-points normalized to $M$ unit cells, and
\begin{align*} 
\chi(z) &= \frac{1}{\pi^{1/4}\sigma^{1/2}} e^{-\frac{1}{2}(\frac{z}{\sigma})^2}  \, 
\end{align*}
is the ground state wave function of the harmonic potential along the $z$-direction with $1/\sqrt{e}$ radius $\sigma$. Evaluating the $z$-integration and using the Bloch functions $\tilde{\psi}_{\pm}$ normalized to a single unit cell leads to
$g_0 = g_{2D} \,I_0$ and $g_1 = g_{2D} \,I_1$ with
\begin{align*} 
g_{2D} &=  \frac{g}{\sqrt{2\pi}\, M A \,\sigma} \, , \\
I_0 &=  A \,\int_{A} dx dy \,|\tilde{\psi}_{\pm}(x,y)|^4 \, , \\
I_1 &=  A \,\int_{A} dx dy \,|\tilde{\psi}_{+}(x,y)|^2 |\tilde{\psi}_{-}(x,y)|^2 \, , \\
\end{align*}
and $g$ denoting the conventional 3D contact interaction strength. The integrals $I_0$, $I_1$ can be derived by an exact band calculation for varying values of $\Delta V_f$.

\begin{figure*}[t!]
\vspace*{1cm}
\includegraphics[width=310pt]{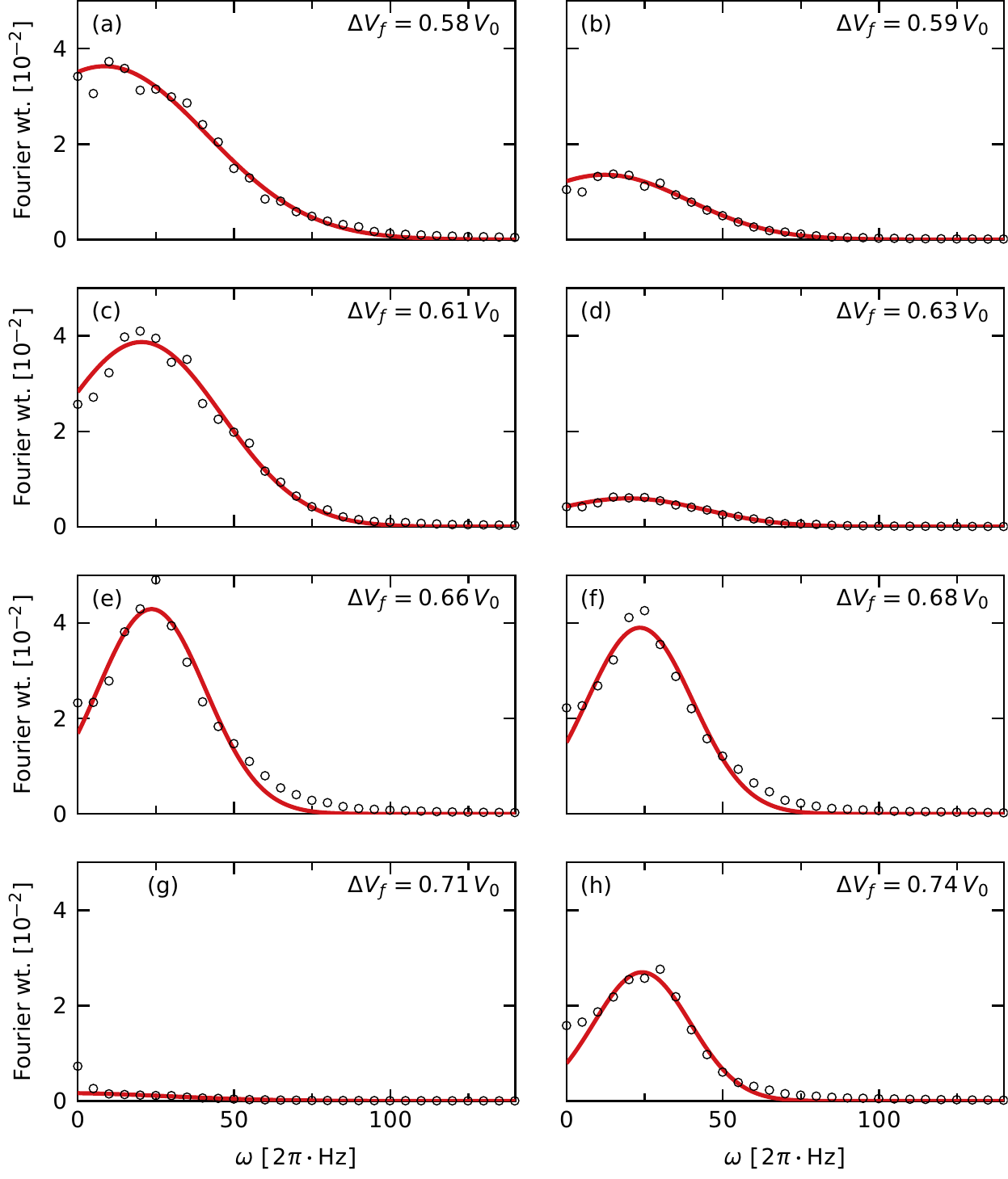}
\caption{Fourier spectrum of the $X$ point oscillations averaged over 500 random initializations (black circles). The peak of the Fourier transform determines the main oscillation frequency at the given value of $\Delta V_f$. The red line shows a Gaussian fit to the Fourier spectrum. We use a temperature $T=76.8\,{\rm nK} \approx 0.8\,E_{\rm rec}/k_B$ comparable to the one used in experiments.}
\label{fig:Fig.A3}
\end{figure*}

\section*{Details on classical-field-theory simulations}
We use a tight-binding model including all nearest- and next-nearest-neighbour hopping terms as well as on-site interaction terms, for details see Ref.~\cite{Nus:20}. Along the $x$- and $y$-direction we use periodic boundary conditions and a discretized harmonic trap in the $z$-direction. We adjust the hopping parameters by optimizing the agreement of the tight-binding band structure to the Bloch band structure and determine the interaction parameters by matching the mean-field interaction strength in the center of each tube to the corresponding estimated experimental value, for details see Ref.~\cite{Nus:20}. We show the resulting set of hopping parameters and on-site potentials in Fig.~\ref{fig:Fig.A2}. Depending on the value of $\Delta V$ the resulting interaction strengths are in the range $0.05 \, E_{\rm rec}< U_A, U_B < 0.1 \, E_{\rm rec}$.

We initialize the system using a Monte-Carlo sampling routine and employ a classical-field-theory simulation for the dynamics. Averaging over several random Monte-Carlo initializations accounts for thermal fluctuations of the initial state. We extract the population of the $X$-points by projecting the wave function obtained from our numerical simulations onto the corresponding tight-binding Bloch functions $\psi_{\pm}$. We proceed similarly for the population of the two interacting lowest-energy states $\psi_{+} \pm i\psi_{-}$.

\begin{figure}[tb]
\includegraphics[width=\linewidth]{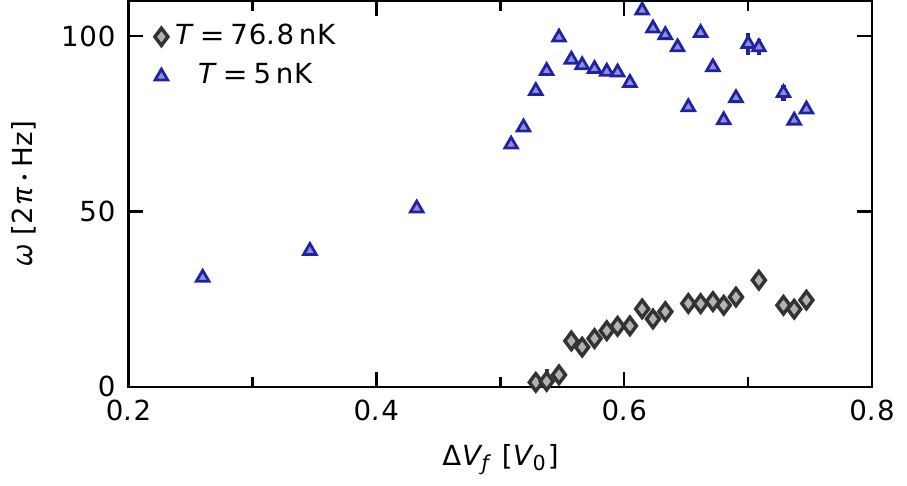}
\caption{Dominant oscillation frequencies obtained from classical-field-theory simulations for several different temperatures as denoted in the legend. For each $\Delta V_f$ we determine the averaged Fourier spectrum, as exemplarily shown in Fig.~3(b) of the main text, and plot the dominant oscillation frequency obtained as the position of the maximum of a Gaussian fit. The error in the determination of this position is smaller than the data symbol. At lower temperature we observe significantly higher oscillation frequencies and smaller FWHM of the Gaussian fit, indicating a sharper peak in frequency space.} 
\label{fig:Fig.A4}
\end{figure}

\section*{Determining the dominant oscillation frequency from Fourier spectra}
Our goal is to extract the dominant frequency component of the oscillation between the two $X$ points. To this end, we Fourier transform the relative $X$ point population $r(t)=\frac{n_+-n_-}{n_++n_-}$ for each individual random Monte-Carlo initialization. For the Fourier transform we use a set of $N$ data points $r_m$ for times $t_m$ with spacing $t_m-t_{m-1}=2.4\,{\rm ms}$ in the range $0=t_{\rm min}<t_i<t_{\rm max}=400\,{\rm ms}$. We use a discrete Fourier transform such that the Fourier weight $F_k$ is
\begin{align*}
F_k&=\left| \frac{1}{N} \sum_{m=0}^{N-1} r_m e^{-2 \pi i \frac{mk}{N}} \right|^2 \quad .
\end{align*}
The Fourier weight $F_k$ corresponds to the frequency $\omega_k=k \Delta \omega$, where $\Delta \omega=1/(t_{\rm max}-t_{\rm min})$. Subsequently we average the Fourier spectrum over many random Monte-Carlo initializations and fit a Gaussian function
\begin{align*}
f(\omega)&= a \exp\left(-\frac{4\, {\rm ln}(2) (\omega-\omega_0)^2}{s_{\rm FWHM}^2}\right)
\end{align*}
to the resulting Fourier spectrum. The peak frequency $\omega_0$ denotes the dominant oscillation frequency. The fitting errors for $\omega_0$ are smaller than the data symbols in Fig.~4 of the main text. We show examples for several Fourier spectra and the corresponding fits in Fig.~\ref{fig:Fig.A3}. 

\section*{Lower temperature}
In figure \ref{fig:Fig.A4} we show the dominant oscillation frequencies for the $X$ point oscillation for $T=76.8\,{\rm nK}$ as in Fig.~4 of the main text and for a 15 times lower initial temperature than in the experiment $T=5\,{\rm nK}$. Since there is less thermal noise at lower temperature the peaks in frequency space are significantly sharper. Additionally the dominant oscillation frequency is shifted to larger frequencies for lower temperatures. We believe that this reflects the correspondingly higher number of atoms condensed at the $X$ points. As a result we can identify the dominant oscillation frequency for a wider range of final potential offsets. The temperature dependance of the oscillation frequencies is found to become significantly weaker at low temperatures. This is a result of the significant energy that is introduced into the system when transferring the atoms to the second band, which dominates the subsequent dynamics rather than the initial temperature before the transfer.  

\begin{figure*}[t!]
\includegraphics[width=420pt]{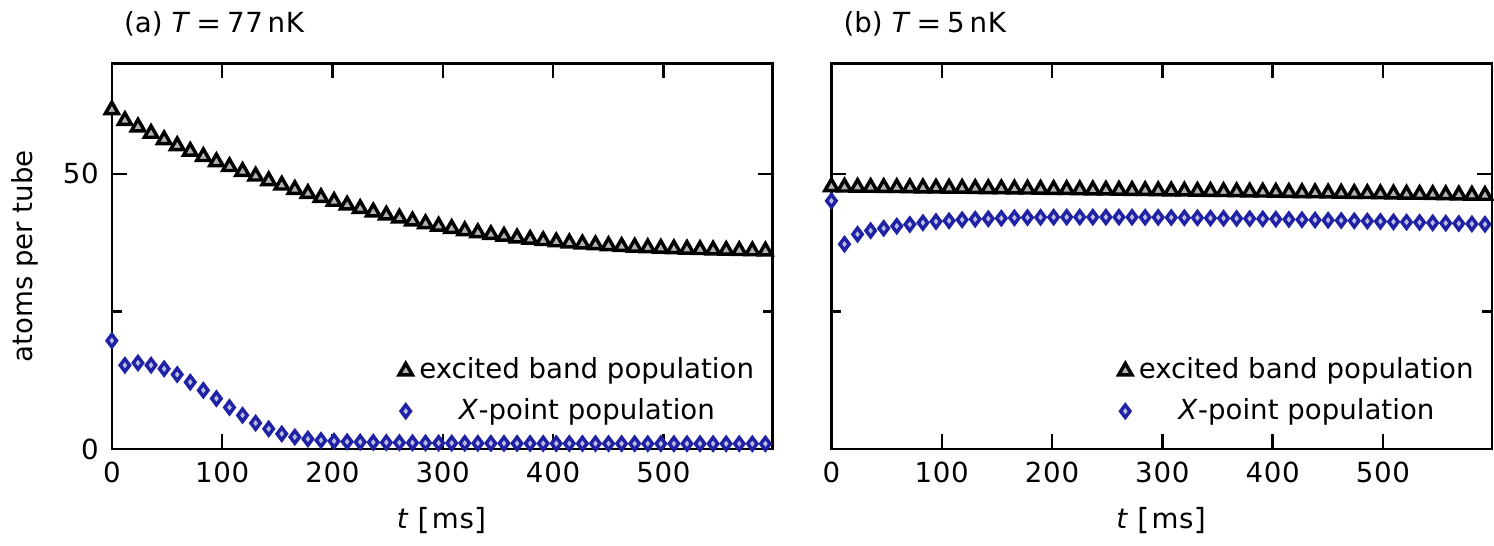}
\caption{Excited band population and corresponding number of atoms at the $X$-points as a function of time. We show the same two cases as in \ref{fig:Fig.A4}. In contrast to the case $T=5\,{\rm nK}$, we see that for $T=76.8\,{\rm nK}$ the population of the $X$-points decays during $100\,$ms in accordance with the decay of the oscillation in Fig.~3(a) of the main text.}
\label{fig:Fig.A5}
\end{figure*}

\section*{Dephasing of oscillations}
\begin{figure*}[b!]
\includegraphics[width=440pt]{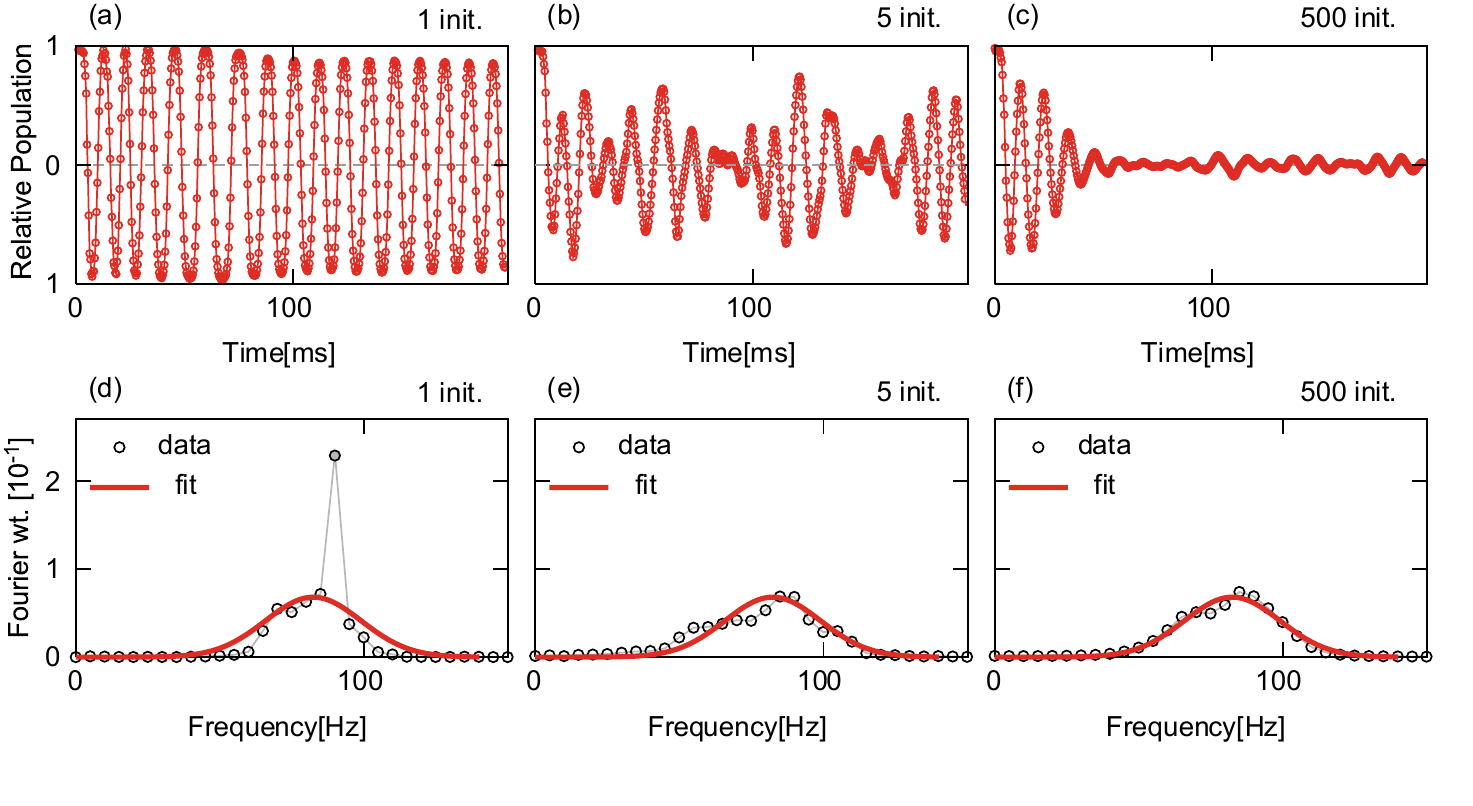}
\caption{(a),(b),(c) Relative population difference of $X$-points $\frac{n_{+}-n_{-}}{n_{+}+n_{-}}$ according to classical-field-theory simulations at $\Delta V_f=0.69\,V_0$. Panel (a) shows the oscillation for a single random initialization, while panels (b) and (c) are averaged over 5 and 500 random initializations, respectively. For each initialization of the system, the oscillation picks a slightly different phase and frequency. Therefore we observe significant damping via dephasing in panels (b) and (c). (d),(e),(f) Fourier spectrum of the oscillation shown in (a) (black circles) averaged over different numbers of random initializations as indicated above the panels. For better comparison we show in all three panels a fit to the average over 500 random initializations (red solid line). For all panels the temperature is $T=5\,{\rm nK}$.}
\label{fig:Fig.A6}
\end{figure*}

In this sub-section, we point out that for temperatures below a few ten nK, i.e. below those realized in the experiment, band decay plays no role. Hence, for sufficiently low temperatures, the origin of the decay of the population oscillations between the $X$-points is dephasing due to different frequency components and not due to loss of coherence or decay to the lowest band. In Fig.~\ref{fig:Fig.A5} we plot the number of atoms in the second band as well as the population at the $X$-points for the same value $\Delta V_f=0.69\,V_0$ and the temperature $T=76.8\,{\rm nK}$ as used in Fig.~3(a) of the main text and for a the far lower temperature $T=5\,{\rm nK}$. In Fig.~\ref{fig:Fig.A5}(a) we see that for $T=76.8\,{\rm nK}$ the population of the $X$-points decays during $100\,$ms in accordance with the decay of the oscillation in Fig.~3(a) of the main text. In contrast, for $T=5\,{\rm nK}$, the second band population and the condensate fractions at the $X$-points remain practically without decay. Hence, we do not expect damping from these two effects. In Fig.~\ref{fig:Fig.A6} we show the oscillations averaged over different numbers of random initializations. While we observe a coherent oscillation for a single initialization, the oscillations are damped significantly for larger numbers of initializations. In the corresponding Fourier spectra we see that a single random initialization has a sharp Fourier peak represented by a single data point in Fig.~\ref{fig:Fig.A6}(d). When averaging over multiple initializations, the Fourier peak becomes broader and for 5 initializations it already acquires a width comparable to the average over 500 initializations. We conclude that damping is a result of dephasing due to oscillations with different frequency components.

\end{appendix}


\begin{thebibliography}{30}
\bibitem{Extrap}
In solving the two-mode model, the number of atoms $N$ determines the dimension of the matrix to be diagonalized. Our numerical capacity has limited $N$ to $5.000$.
\end{thebibliography}
\end{document}